\documentclass[aps,prl, twocolumn, showpacs, superscriptaddress,preprintnumbers, amsmath, epsfig, floatfix,normalem]{revtex4-1}

\usepackage{graphicx}
\usepackage{physics}
\usepackage{dcolumn}
\usepackage{bm}
\usepackage{ulem}
\usepackage{subfigure}
\usepackage[utf8]{inputenc}
\usepackage[english]{babel}
\graphicspath{{figs/}{figs:}{\figs}}
\usepackage[usenames,dvipsnames]{color}
\usepackage[]{ulem}
\usepackage[colorlinks=true,
linkcolor=blue,
urlcolor=blue,
citecolor=blue,
breaklinks=true]{hyperref}
\newcommand{\comment}[1]{\textcolor{red}{#1}}
\renewcommand{\comment}[1]{\relax}
\newcommand{\todelete}[1]{\textcolor{green}{\sout{#1}}}
\renewcommand{\todelete}[1]{\relax}

\newcommand*{\balancecolsandclearpage}{%
\close@column@grid
\clearpage
\twocolumngrid}

\begin{document}

\title{Parametrically excited star-shaped patterns at the interface\\ of binary Bose-Einstein condensates}
\date{\today}

\author{D. K. Maity}
\affiliation{Department of Physics, Indian Institute of Technology Kharagpur, Kharagpur-721302, India}
\author{K. Mukherjee}
\affiliation{Department of Physics, Indian Institute of Technology Kharagpur, Kharagpur-721302, India}
\affiliation{Center for Optical Quantum Technologies, Department of Physics, University of Hamburg, Luruper Chaussee 149, 22761 Hamburg Germany}
\author{S. I. Mistakidis}
\affiliation{Center for Optical Quantum Technologies, Department of Physics, University of Hamburg, Luruper Chaussee 149, 22761 Hamburg Germany}
\author{S. Das}
\affiliation{Department of Physics, Indian Institute of Technology Kharagpur, Kharagpur-721302, India}
\author{\\P. G. Kevrekidis}
\affiliation{Department of Mathematics and Statistics, University of Massachusetts Amherst, Amherst, MA 01003-4515, USA} \affiliation{Mathematical Institute, University of Oxford, OX26GG, UK}
\author{S. Majumder}
\affiliation{Department of Physics, Indian Institute of Technology Kharagpur, Kharagpur-721302, India}
\author{P. Schmelcher}
\affiliation{Center for Optical Quantum Technologies, Department of Physics, University of Hamburg, Luruper Chaussee 149, 22761 Hamburg Germany}\affiliation{The Hamburg Centre for Ultrafast Imaging, University of Hamburg, Luruper Chaussee 149, 22761 Hamburg, Germany}
	
\begin{abstract}
\noindent
A Faraday-wave-like parametric instability is investigated via mean-field and Floquet analysis in immiscible binary Bose-Einstein condensates. 
The condensates form a so-called \textit{ball-shell} structure in a two-dimensional harmonic trap. 
To trigger the dynamics, the scattering length of the core condensate is periodically modulated in time. 
We reveal that in the dynamics the interface becomes unstable towards the formation of oscillating patterns. 
The interface oscillates sub-harmonically exhibiting an $m$-fold rotational symmetry that can be controlled by maneuvering the amplitude and the frequency of the modulation. 
Using Floquet analysis we are able to predict the generated interfacial tension of the mixture and derive a dispersion relation for the natural frequencies of the emergent patterns. 
A heteronuclear system composed of $^{87}$Rb-$^{85}$Rb atoms can be used for the experimental realization of the phenomenon, yet our results are independent of the specifics of the employed atomic species {and of the parameter at which the driving is applied.}   
\end{abstract}
\maketitle 

{\it Introduction.--} Liquid drops or puddles~\cite{Noblin, Brunet} that are weakly affixed to a vertically oscillating surface, or a periodically driven spherical liquid drop levitated from the surface, either acoustically~\cite{Shen} or magnetically~\cite{Hill}, can display star-shaped patterns. 
These patterns constitute a paradigm of a spatial as well as temporal symmetry breaking phenomenon. Their appearance in the form of the so-called Faraday-pattern dates back to 1831 for a fluid in a vertically shaken vessel~\cite{Faraday_1831}. 
In close resemblance to the original Faraday-experiment, the symmetry-breaking instabilities have been intensively studied in classical fluids considering a variety of surfaces of the liquid such as spherical~\cite{Adou_Tuckerman_2016, Li}, cylindrical~\cite{Dilip,Dilip_2} and flat~\cite{edwards_fauve_1994, Kumar_1996}. 
Remarkably, the dominant wavelength of the instability, and the symmetries of the emergent patterns are determined by a few intrinsic properties such as the density and the surface tension of the liquid~\cite{Takaki_Adachi, douady_1990}.

Over the last two decades, Bose-Einstein condensates (BECs) due to their remarkable experimental tunability~\cite{Chin2010, Thorsten2006,bloch2008many} have facilitated the investigation of various classical hydrodynamical instabilities in the context of quantum fluids. 
Indeed, several theoretical works unveiled the emergence of
parametric resonances~\cite{Pedro1999} and Faraday waves~\cite{Staliunas, Nath, Antun2014}, either via confinement modulations~\cite{Modugno,Nicolin} or by means of a
time-dependent scattering length~\cite{Staliunas}, inspiring also the experimental realization of parametric resonances~\cite{Pollack} and Faraday-waves in BECs ~\cite{Engels,  Nguyen}.

Importantly, even though such modulation dynamics has been extensively unraveled for single-component BECs~\cite{Pedro1999, ueda02, abdu03, pgk03, Staliunas,Staliunas2004, Kumar2008, Nicolin, Nicolin2011, Chen2018}, the corresponding two-dimensional (2D) multispecies BEC scenario is far less explored~\cite{nico12,nath12,abdu13,Li2016, Chen2019}. 
In this context, multispecies BECs exhibiting a well-defined interface~\cite{Trippenbach_2000,Barankov2002, Lee2016, Indekeu2015} can emulate some of the well-known interfacial tension dominated fluid instabilities. 
These include the Rayleigh-Taylor~\cite{Sasaki,Gautam,Kadokura}, the capillary~\cite{Sasaki_K, Indekeu2018}, the Kelvin-Helmholtz~\cite{Takeuchi,Suzuki,Baggaley}, the Richtmyer-Meshkov~\cite{Bezett}, the countersuperflow~\cite{Law,Yukalov,Takeuchi_2,Hamner}, and the Rosensweig instability~\cite{Saito} as well as the Benard-Von-Karman vortex street~\cite{Sasaki_2,Sasaki_3}. 
Recall that for multispecies BECs the interfacial tension stems from the combined effect of quantum pressure and interspecies interactions~\cite{Van2008}. 
Interestingly, harmonically trapped binary condensates in quasi-2D can form
a circular interface between them, rendering these systems ideal candidates to probe azimuthal-symmetry breaking instabilities. 
The analogues of the latter in classical fluids are extremely useful in experiments~\cite{Noblin, Brunet, douady_1990} for determining the surface tension of the liquid. 
Therefore, it would be extremely desirable to explore whether interfacial pattern formation in 2D binary BECs can provide information regarding the interfacial tension of the mixture.

\begin{figure}[ht]
\begin{center}
\includegraphics[height=!, width=0.47\textwidth]{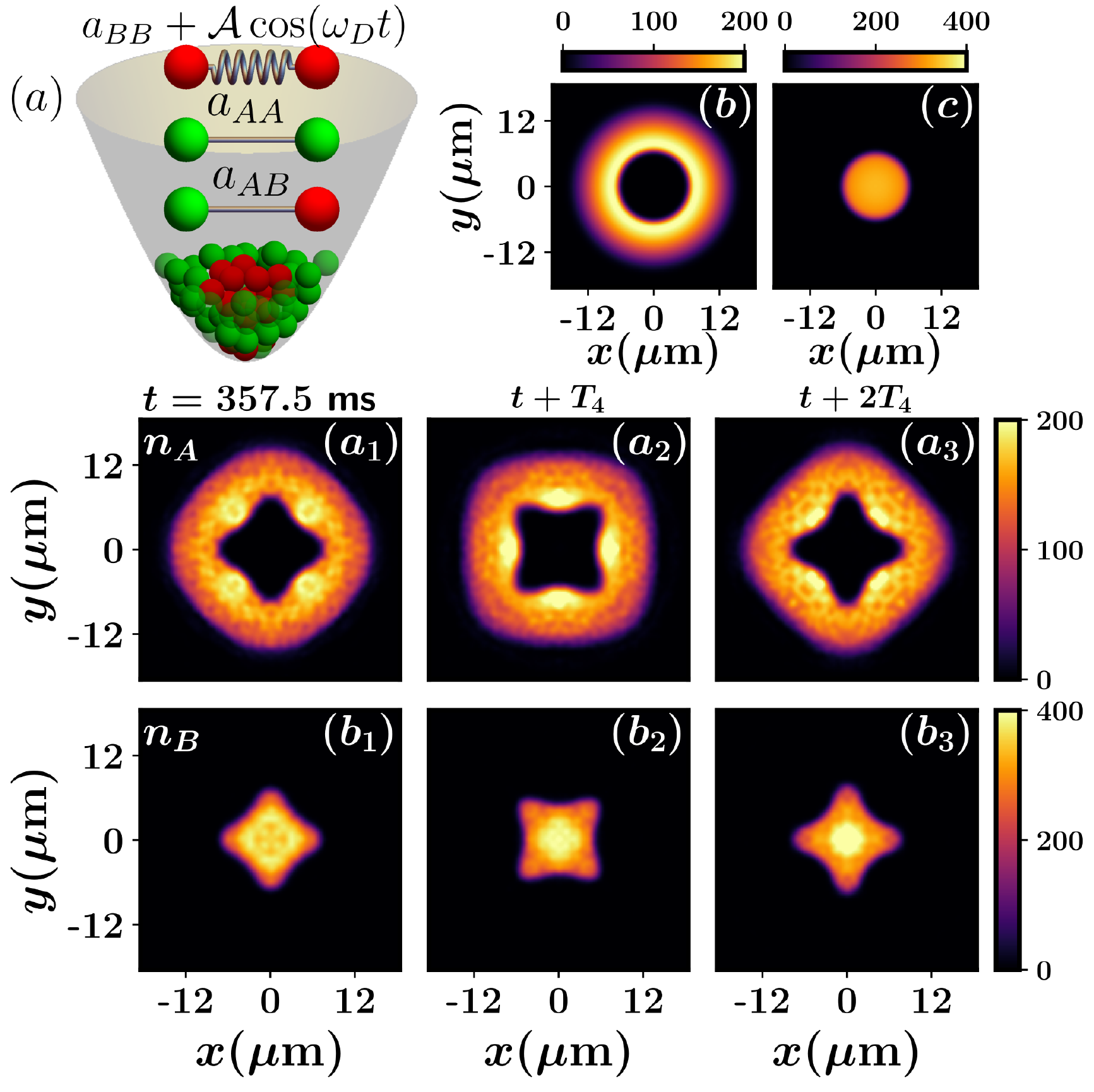}
\caption{(Color online) (a) Schematic representation of the 2D harmonically trapped binary BEC and the modulation protocol. 
Density profiles of ($b$) species A and ($c$) species B at the initial state where the intra- and the interspecies scattering lengths are $a_{AA} = 99a_{B}$, $a_{BB} = 75a_B$ and $a_{AB} = 213a_B$. Snapshots of the density profiles of ($a_1$)-($a_3$) species A and ($b_1$)-($b_3$) species B at consecutive periods $nT_4$ of the modulation (see legend). 
The $^{87}$Rb-$^{85}$Rb binary BEC consisting of $N_A=10^4$ and $N_B=5\times 10^3$ atoms is prepared as in Figs.~\ref{fig2} ($b$), ($c$). 
The dynamics is initiated following a periodic modulation of $a_{BB}$ with frequency $\omega_D/2\pi = 1/T_4 = 69$ Hz and amplitude $\mathcal{A} = 15 a_B$. 
The star-shaped pattern $D_4$ exhibits four lobes, see also videos of the dynamics in \cite{Supplementary}.}
\label{fig2}
\end{center}
\end{figure}

In this letter, we propose  a parametrically driven
mechanism that enables on-demand azimuthal-mode
pattern formation at the interface between two immiscible BECs.
This, on the one hand, lends further support to the striking similarity between
of a number of features shared between the classical fluid and the BEC systems and, on the other hand, paves the way to determine the interfacial tension in ultracold atom experiments based on the resulting patterns occurring at the interface among the components. 
More specifically, we consider a binary BEC of two different atomic species confined in a 2D axisymmetric harmonic trap (see Fig.~\ref{fig2}($a$)) and interacting via short-range repulsive interactions. 
The mixture is initialized in a radially symmetric, phase-separated configuration \cite{mertes2007nonequilibrium, Papp}, where the species with weaker interactions is surrounded by the one with stronger ones, thus forming a so-called {\it ball-shell} structure with  a circularly symmetric interface [Figs.~\ref{fig2}($b$), ($c$)]. 
Upon applying a time-periodic modulation of the intraspecies core component interaction we demonstrate that both the spatial and the temporal symmetries of the interface are broken,
for a generalization of the results to other dynamical protocols see~\cite{supmat}. 
Importantly, depending on the modulation strength and frequency, star-shaped density patterns $D_m$, with underlying $m$-fold rotational symmetry, appear at the interface which oscillates sub-harmonically, i.e. at half of the modulation frequency. 
To elucidate the emergence of the resulting patterns, the underlying mean-field equations are reduced to a Mathieu equation at the level of the amplitude of the $m$-th mode where Floquet theory is subsequently applied. 
A dispersion relation relating the azimuthal wavenumber $m$ of the pattern to its frequency $\omega_m$ is derived. 
The stability boundaries of the resulting star-shaped
patterns are identified at the level of the full mean-field model being in good agreement with the predictions of the effective theoretical analysis in terms of the Floquet theory. 
Remarkably, it is demonstrated that this dispersion relation can be employed to predict the interfacial tension of the phase-separated BEC. 

{\it Model.--} Our theoretical approach to describe the dynamics of the binary BEC relies on the coupled system of time-dependent Gross-Pitaevskii (GP) equations~\cite{pethick,string} 
\begin{equation}\label{GP}
\begin{split}
& i\hbar\frac{\partial }{\partial t}\Psi_j(\vb{r}, t) =  \bigg\{ -\frac{\hbar^2}{2m_j} \bigg [\frac{1}{r}\frac{\partial}{\partial r} \bigg (r\frac{\partial}{\partial r} \bigg)  + \frac{1}{r^2}\frac{\partial^2}{\partial \theta^2} + \frac{\partial^2}{\partial z^2} \bigg ] \\ & + \frac{1}{2}m_j \omega^2_j \big( r^2 + \lambda_j z^2 \big) +
g_{jj}\abs{\Psi_j}^2 + g_{jj^{'}} |\Psi_{j^{'}}|^2\bigg\} \Psi_j(\vb{r}, t).
\end{split}
\end{equation}
Here, {$\vb{r} \equiv ( r,\theta, z )$ denotes the cylindrical coordinates,} $j, j' \in (A, B)$, and the wavefunction $\Psi_j$ of species-$j$ satisfies $\int {|\Psi_j|}^2 \boldsymbol{dr}=N_j$. 
Also, $N_j$, $m_j$ and $\omega_j$ denote the atom number, the mass, and the transverse harmonic trap frequency of species-$j$ respectively. 
{The parameter $\lambda_j$ is the ratio of the axial and the transverse trap frequencies of species-$j$ satisfying here $\lambda_j = \omega_{zj}/\omega_j\gg 1$ which ensures that the dynamics in the $z$-direction is ``frozen-out''.}
The intra- and interspecies interaction strengths $g_{AA}$, $g_{BB}$ and $g_{AB}$ satisfy the so-called phase-separation condition $g_{AB}g_{BA}/(g_{AA} g_{BB}) \propto a^2_{AB}/(a_{AA} a_{BB}) \geq 1$~\cite{Ao1998, Timmermans1998}, where $a_{jj}$ and $a_{jj'}$ are the corresponding $s$-wave scattering lengths. 
\begin{figure*}[ht]
\begin{center} 
\includegraphics[height = 0.2\textheight, width = \linewidth]{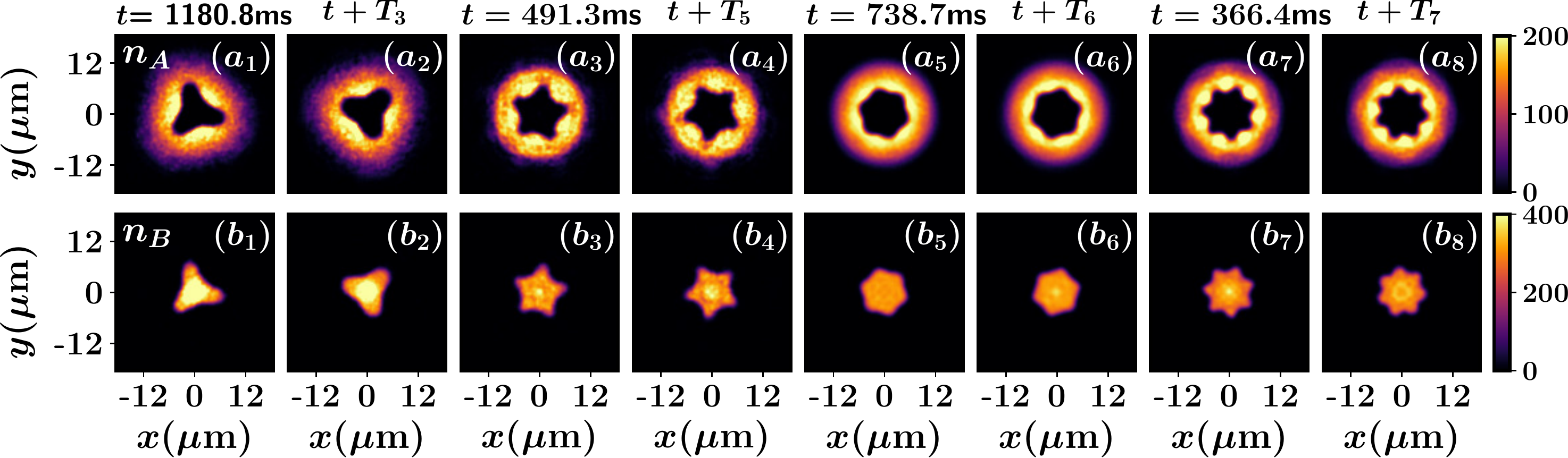}
\caption{(Color online) Density profiles of ($a_1$)-($a_8$) species A and ($b_1$)-($b_8$) species B for various symmetric patterns $D_3$ [($a_1$), ($a_2$), ($b_1$), ($b_2$)], $D_5$ [($a_3$), ($a_4$), ($b_3$), ($b_4$)], $D_6$ [($a_5$), ($a_6$), ($b_5$), ($b_6$)] and $D_7$ [($a_7$), ($a_8$), ($b_7$), ($b_8$)] at specific time-instants (see legends) in the long-time dynamics. 
$T_m$ is the driving period corresponding to the $m$-fold symmetric pattern $D_m$ which is generated by periodically modulating the scattering length $a_{BB}$ in time with amplitude $\mathcal{A} = 15a_B$ and frequency $\omega_D = 2 \pi/T_m$, see also \cite{Supplementary} for videos of the dynamics. 
The $^{87}$Rb-$^{85}$Rb binary BEC consists of $N_A=10^4$ and $N_B=5\times 10^3$ atoms initially prepared as in Fig.~\ref{fig2}.}
\label{fig3}
\end{center}
\end{figure*}
Equation~\eqref{GP} is solved using a split-time Crank-Nicolson method~\cite{Muruganandam2009, VUDRA_2021} in imaginary-time to obtain the initial ground state, and in real-time to monitor the dynamics, in 2D, characterized by the wavefunctions $\psi_j(r,  \theta)$. For the dynamical reduction from 3D to 2D, see~\cite{supmat}. 
With the initial state at hand, we trigger the interfacial dynamics of the binary BEC by periodically modulating the scattering length $a_{BB}$ according to $\tilde {a}_{BB}= a_{BB}+ \mathcal{A}\cos(\omega_D t)$, where $\mathcal{A}$ and $\omega_D$ are the amplitude and frequency of the modulation. 

The experimentally relevant parameters of a $^{87}$Rb-$^{85}$Rb binary BEC labeled as species A and species B are utilized, namely $\omega_A/2\pi = \omega_B/2 \pi \equiv \omega/2 \pi = 15$ Hz, $N_A = 10^4$, $N_B = 5 \times 10^3$, $a_{AA} = 99a_B$ and $a_{AB} = 213 a_B$, with $a_B$ being the Bohr radius~\cite{Papp}.
Accordingly, the initial state corresponds to a shell-structured geometry in which $^{85}$Rb atoms occupy the central region of the trap [Fig.~\ref{fig2}($c$)], hence referred to as the core-condensate, while $^{87}$Rb atoms form a lower density shell [Fig.~\ref{fig2}($b$)] surrounding the core-condensate. 
Since the scattering length $a_{BB}$ can be experimentally tuned via a Feshbach resonance in the range of $50a_B$-$900a_B$~\cite{Papp}, we periodically vary $a_{BB}$ in time around $a_{BB} =75 a_B$ which ensures that the phase-separated condition is fulfilled throughout the dynamics.
Also as a case example we use a modulation amplitude $\mathcal{A} = 15a_B$, for variations of this parameter see also the discussion below. 
This driving process leads, in the long-time dynamics, to the formation of patterns at the interface which steadily oscillates at half of the modulation frequency. 
{Importantly, the observed pattern formation occurs also for different atomic species than the ones considered herein or hyperfine states of the same isotope and irrespectively of the periodic driving of the involved intra- or interspecies scattering lengths, i.e. it represents a {\it generic} phenomenology of the immiscible two-component system, see  \cite{supmat}.}

{\it Results and Discussion.--} Representative density profiles of each species, $n_j = \abs{\psi_j}^2$, unveiling the dynamical generation of a star-shaped pattern $D_4$ with $m =4$ lobes at time-instants representing
integer multiples of the modulation period $T_4$ are illustrated in Fig.~\ref{fig2} following a periodic oscillation of $a_{BB}$ with $\omega_D/2\pi= 69$ Hz. 
It becomes apparent that the four-lobed star pattern [Figs.~\ref{fig2}($a_1$), ($b_1$)] dynamically appears at the interface at $t=357.5$ ms for the first time. 
Indeed, the instability in the system grows until it is clearly visible in the density profiles after about $357.5$ ms. 
The exactly same structure re-appears at time $t + 2T_4$ [Figs.~\ref{fig2}($a_3$) and ($b_3$)], thus revealing its sub-harmonic nature. 
Note that the lobes of the $D_4$ pattern at $t = T_4$ are oriented in
a way rotated by an angle $\pi/4$ with respect to the one at $t = 357.5$ ms or $ t + 2T_4$ [Figs.~\ref{fig2}($a_2$), ($b_2$)]. 
Importantly, patterns with higher $m$-fold rotational symmetries can also be dynamically generated. 
To achieve this we fix the modulation amplitude at $\mathcal{A} = 15 a_{B}$, and change the modulation frequency $\omega_D$. 
As we shall explain later on, a certain symmetric pattern $D_m$ is realized within a specific interval of $\omega_D$ for a given amplitude $\mathcal{A}$, while outside this interval the pattern disappears. 
Prototypical examples of relevant density profiles for symmetric patterns $D_3$, $D_5$, $D_6$ and $D_7$ with three, five, six and seven lobes realized at modulation frequencies $\omega_D\approx48$ Hz, $\omega_D\approx95.1$ Hz, $\omega_D\approx132$ Hz, and $\omega_D\approx157.5$ Hz respectively are depicted in Fig.~\ref{fig3}. 
Indeed, the same density pattern $D_m$ is repeated, but with a different orientation (at an angle $\pi/m$) compared to the earlier one, after every single period of the modulation. 

To expose the nature of the interfacial dynamics, we next perform a linear stability analysis based on the Floquet technique~\cite{Kumar_1996, Staliunas2002,goldman2014periodically,barone1977floquet,eckardt2017colloquium}, by assuming that both species possess a uniform density with a sharp boundary between them. 
Indeed, within the GP calculations we observe that the BEC background density undergoes only a small amplitude breathing motion. 
This allows us to neglect the effect of local density fluctuations within the stability analysis. 
We further presume that the same interface dynamics can be retrieved following the periodic modulation of any other prototypical system parameter (i.e., the different scattering lengths or trap strengths), since the natural angular frequencies of the interfacial patterns should be independent of the parameter at which the periodic protocol is applied. 
To this end, for the convenience of the theoretical analysis a time-dependent harmonic potential of the core condensate is considered instead of the periodic driving of its scattering length, see also \cite{supmat}. 
The adjustable nature of both the trapping potential and the scattering length by means of a tunable magnetic field in typical BEC experiments \cite{inouye2004observation,Chin2010,grimm2000optical} motivates further this assumption, at least, for predicting the natural frequencies of the interface patterns. 

For simplicity we again start our analysis in 3D and subsequently reduce the problem to 2D. 
It is appropriate to express the condensate wavefunction according to the Madelung transformation~\cite{madelung1927quantentheorie} i.e. $\Psi_j(r, \theta, z) = \sqrt{n_j(r, \theta, z)}e^{i \phi_j }$, where $n_j$ and $\phi_j$ are the density and the phase of species-$j$. 
Correspondingly, the superfluid velocity $\boldsymbol{v}_j=(v_{jr}, v_{j\theta}, v_{jz})$ is defined as $\boldsymbol{v}_j = \frac{\hbar}{m_j}\nabla \phi_j$, where $\phi_j$ is also known as the velocity potential~\cite{pethick_smith_2008}. 
Furthermore, it is reasonable to approximate $n_A = 0$ for $r < R$ and $n_B = 0$ for $r > R$ with $R$ being the radius of the interface. 
Therefore, the coupled GP system of Eq.~\eqref{GP} can be expressed in the form~\cite{pethick_smith_2008}
\begin{equation}\label{velocity}
- m_j \frac{\partial \boldsymbol{v}_j}{\partial t} = \frac{\vb{\nabla} P_j}{n_j},~~~~~\nabla^2 \phi_j = 0.
\end{equation}
The effective pressure term of the individual species is $P_A = \frac{1}{2}(m_A n_A v^2_A) + \frac{\hbar^2 \sqrt{ n_A}}{2 m_A } \nabla^2 \sqrt{n_A} + g_{AA}n^2_A + \frac{1}{2}m_A n^2_A \omega^2 (r^2 + \lambda_A z^2) $ and $P_B = \frac{1}{2}(m_B n_B v^2_B) + \frac{\hbar^2 \sqrt{n_B}}{2 m_B } \nabla^2 \sqrt{n_B} + g_{BB}n^2_B + \frac{1}{2}m_B  n_B \omega^2 ( r^2 + \lambda_B z^2 ) + \frac{1}{2}m_B  n_B \omega^2 r^2 b \cos(\omega_D t))$, respectively. 
Note that the problem has been effectively reduced to two single-component
problems interacting through their sharp interface. 
The influence of $g_{AB}$ is still implicitly incorporated in the values of $R$, $n_B$ and $n_A$ taken from the full GP model. 
Importantly, here, the amplitude $b$ is related to $\mathcal{A}$ (in the GP level) by $b = 2\mathcal{A} \abs{\psi_B(r = R)}^2/(m_B \omega^2 R^2)$~\cite{supmat}. 
This relation is derived in order to have the same dynamical impact on the interface by both protocols~\cite{supmat}. 
After the onset of the instability, the interface is deformed by a small amount $\zeta$. 
The stress balance condition (also known as Laplace’s formula in fluid mechanics) at the interface can be written as ~\cite{Lamb_1932, Sasaki_2011}
\begin{equation}\label{SB}
\bigg[ P_{B}-P_{A} \bigg ]_{r=R+\zeta}={\sigma}\left[{\frac{1}{R_1}+\frac{1}{R_2}}\right],
\end{equation}
$R_1$, $R_2$ are the principal radii of the interface curvature and $\sigma$ denotes the interfacial tension. 
Linearizing Eq.~\eqref{SB} around $r = R$, and using Eq.~\eqref{velocity} and the kinematic boundary condition~\cite{Lamb_1932} i.e. $
\frac{\partial \zeta}{\partial t}=v_{Ar}(r=R)=v_{Br}(r=R)$, and considering $k\to 0$ (no-excitation along the $z$-direction) we arrive at a Mathieu-type equation~\cite{supmat} for $\zeta_m$ namely  
\begin{equation}\label{mathieu}
\ddot{\zeta}_m+\omega_{m}^2 \big[ 1 - ( b/b_{0m}) \cos{(\omega_D t)} \big]\zeta_m =0,
\end{equation}
with $\omega_m^2=\frac{\sigma}{R^3} \frac{m(m^2-1)}{(m_B n_B - m_A n_A)}$ and $b_{0m}=\frac{\sigma(m^2-1)}{ m_B \omega^2 n_B R^3}$.
\begin{figure}
\begin{center}
\includegraphics[height=!, width=8.2 cm]{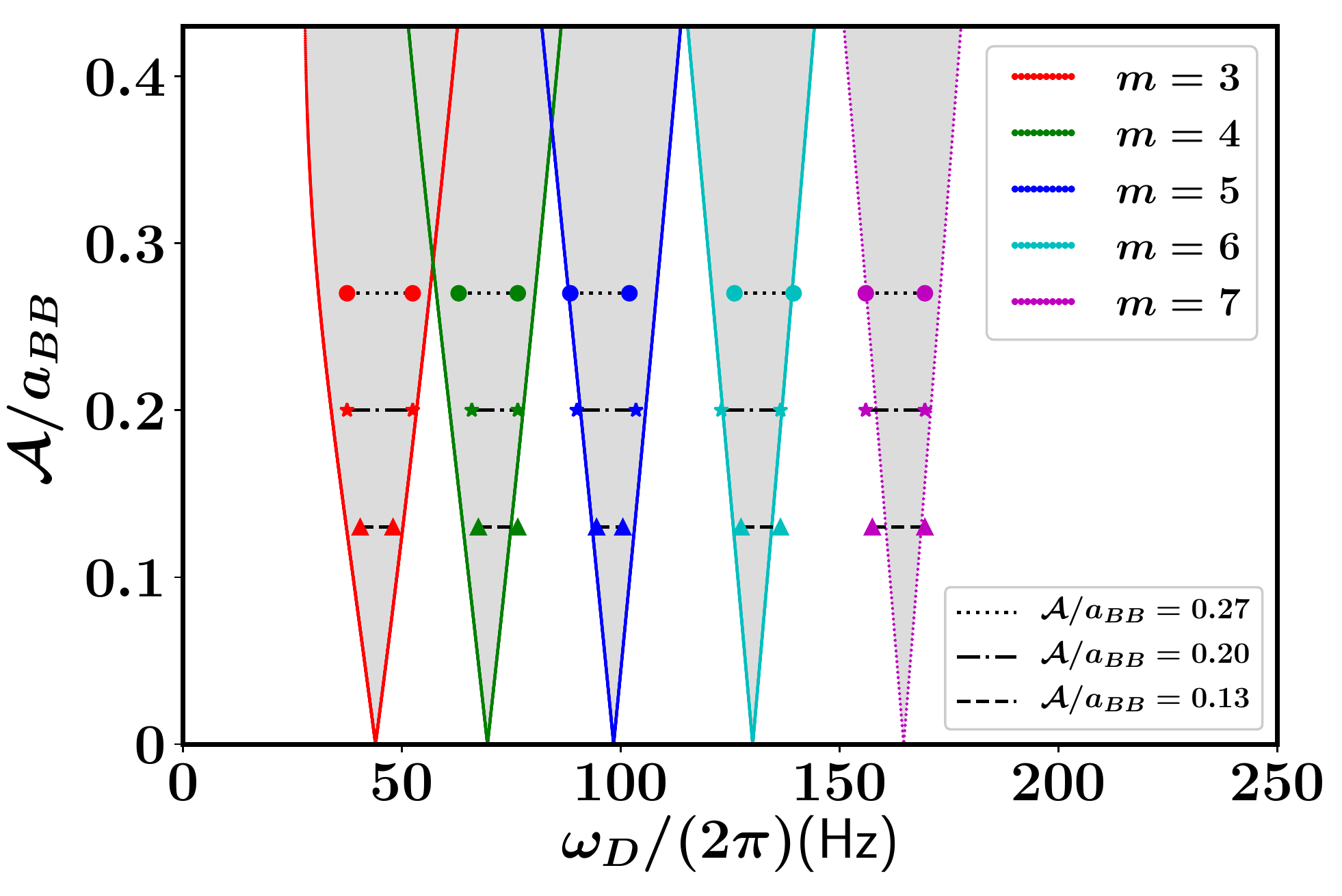}
\caption{(Color online) The marginal stability boundaries, predicted within Floquet theory, for the first sub-harmonic excitations of different azimuthal wavenumbers $m$. 
The system is unstable inside the tongues (shaded regions) towards the formation of $m$-fold star-shaped symmetric patterns $D_m$. 
Two starlike (circular) [triangle] data points connected by a dashed (dotted) [dashed-dotted] line within each tongue indicate the frequency interval for the fixed amplitude $\mathcal{A} = 0.20a_{BB}$ ($0.27a_{BB}$) [$0.13a_{BB}$] where GP calculations have been performed.}
\label{fig4}
\end{center}
\end{figure}
Remarkably, the expression of the natural frequency $\omega_m$ has the same form as the one of a classical inviscid incompressible fluid~\cite{Takaki_Adachi, Strutt1879}. 
According to Floquet theory, there exists a solution $\zeta_m (t) = e^{(s+i\alpha \omega_D ) t} \sum_{p=-\infty}^{\infty} \zeta_m^{(p)} e^{i p\omega_{D} t}$, where $s$ is the growth rate and $\alpha$ is the Floquet exponent. Inserting this into Eq.~\eqref{mathieu}, the Floquet expansion of $\zeta_m$ leads to a linear difference equation
\begin{equation}\label{differnce}
A_m^{(p)}\zeta_m^{(p)} = b \left( \zeta_m^{(p-1)}+\zeta_m^{(p+1)}\right),
\end{equation}
with $A_m^{(p)} = \left[2(p+\alpha)^2 \omega^2_D \frac{(m_B n_B - m_A n_A)}{mm_B \omega^2 n_B}-\frac{2\sigma(m^2-1)}{R^3 n_B m_B \omega^2 }\right]$. 

The eigenvalues $b(\omega_D, m)$ of Eq.~\eqref{differnce}, depending on the parameter $s$, describe the stability of the system in the parameter space of the driving amplitude $\mathcal{A}$ and frequency $\omega_D$. 
Namely, the system is unstable when $s>0$ and pattern formation is expected to occur at the interface. In particular, we let $s=0$ which provides the marginal stability boundaries for different values of the azimuthal wavenumber $m$, see Fig.~\ref{fig4}. 
Recall that for sub-harmonically excited waves the relation $\omega_m = \omega_D/2$ holds, which allows us to set $\alpha = 1/2$. 
In principle, the stability curve is composed of an infinite series of resonant tongues. 
However, in Fig.~\ref{fig4} we showcase the first sub-harmonic marginal stability boundaries for $m=3,4,5,6,7$ since those are the ones that we have shown previously in the form of star-shaped patterns in the GP framework. 
Indeed, the system will be unstable exhibiting pattern formation if $\mathcal{A}$, $\omega_{D}$ lie inside the boundaries of a specific tongue [see e.g. the stars in Fig.~\ref{fig4}] but it is stable when they reside outside these tongues. 

To expose the reliability of the predictions of Floquet theory we also present in Fig. \ref{fig4} with data points the borders at which pattern formation in terms of $\omega_D$, $\mathcal{A}$ occurs within the GP theory. 
For instance, we find that for $0.13<\mathcal{A}<0.27$ all different patterns can be realized within the mean-field framework meaning that only some of the $m$-fold structures cannot be captured for other amplitudes. 
Another key observation here is that for a fixed $\mathcal{A}$ the widths of the stability boundaries in the effective theory appear to decrease for larger $\omega_D$ but they show a non-monotonic behavior in the mean-field calculations [Fig.~\ref{fig4}]. 
This deviation (as well as similar non-monotonicities for fixed $\omega_D$ and varying $\mathcal{A}$) can be attributed to the linear nature of the theoretical model ignoring possible non-linear effects captured in the GP framework. 
Note that the symmetries of the patterns depend crucially on $\omega_D$, i.e. for a larger $\omega_D$ higher-fold symmetries appear in the system. 
Furthermore, recall that the patterns are repeated after $t+nT_m$, $n=1,2,\dots$, being rotated by $n\pi/m$ with respect to the one at $t$. 
This behavior is associated with the spatial and temporal symmetry of the system which can be explained from the Floquet analysis. 
For instance, $\zeta(\theta,t)=\sum_{m=1}^{\infty}{\zeta}_m(t) e^{i{m\theta}}$ is invariant under the transformations $\theta \rightarrow \theta + (n\pi/m)$ and $t \rightarrow t  + nT$, which exactly correspond to the spatial and temporal reprisal of the patterns according to the predictions of the GP calculations. 
Finally, in order to calculate the interfacial tension, we compare our GP results ($\mathcal{A}$, $\omega_{D}$) for a particular symmetric pattern to the instability tongues emerging from the Floquet analysis [Eq. (\ref{mathieu})]. 
This gives a value of the interfacial tension $\sigma=1.1 \times 10^{-18}\pm 5 \% $ N/m \cite{error}.

{\it Conclusions/Future Challenges.--} The interface dynamics of a phase-separated binary BEC subjected to a periodic modulation of the core component scattering length has been investigated using mean-field theory and a Floquet analysis. 
We have found that, depending on the driving frequency, the interface becomes unstable to azimuthal undulations.
As a result of the instability, it is possible to controllably induce
patterns of $m\geq 3$-fold rotational symmetry on the immiscible
two-component system and to
predict their symmetries, as well as their subsequent time evolution
and recurrence. 
Utilizing Floquet analysis we derived a dispersion relation which allows
us to predict the natural frequencies of the emergent patterns. 
Most importantly, in cold-atom experiments, these patterns and the corresponding driving frequencies, can be employed to determine the interfacial tension. 
In the realm of two spatial dimensions, it would be intriguing to examine the corresponding instabilities and consequent pattern formation in the presence of dipolar,
as well as spin-orbit interactions. 
Certainly, deriving an effective quasi-1d equation~\cite{wenlong} characterizing the interface dynamical evolution, beyond the linearized stage considered herein, would be an interesting perspective.  
Lastly, to connect with experiments such as the one
of~\cite{mertes2007nonequilibrium}, it would be relevant to extend considerations
to a fully 3D setting.
	
\textit{Acknowledgments} 
K.M. acknowledges a research fellowship (Funding ID no 57381333)  from  the Deutscher Akademischer Austauschdienst (DAAD). 
S.I.M. gratefully acknowledges financial support in the framework of the Lenz-Ising Award 
of the University of Hamburg. 
K.M thanks A.K Mukhopadhyay for a careful reading of the manuscript and insightful discussions.
This material is based upon work supported by the US National Science
Foundation under Grants No. PHY-1602994 and DMS-1809074
(PGK). PGK also acknowledges support from the Leverhulme Trust via a
Visiting Fellowship and thanks the Mathematical Institute of the University
of Oxford for its hospitality during part of this work.

\bibliographystyle{apsrev4-1.bst}
\bibliography{reference.bib}{}

\newpage
 
\pagebreak
\widetext
\clearpage

\begin{center}
\textbf{\large Supplemental Material: Parametrically excited star-shaped patterns at the interface\\ of binary Bose-Einstein condensates}
\end{center}

\twocolumngrid

\section{Dimensional reduction and Computational Details}

Let us elaborate on how the full three-dimensional (3D) Gross-Pitaevskii (GP) equations of motion boil down to their two-dimensional (2D) form used for the calculations presented in the main text. The coupled set of full 3D GP equations, discussed in the main text, can be cast into a dimensionless form by scaling the spatial coordinates as $x' = x/a_{\rm osc}$, $y' =y/a_{\rm osc}$, $z' = z/a_{\rm osc}$, the time as $t'= t/\omega_A$, and the wavefunction as $\Psi'_{j}(x', y', z') = \sqrt{a_{\rm osc}^3/N_{j}} \Psi_{j}(x, y, z, t)$. 
Here, the index $j=A,B$ refers to each of the species of the binary BEC, while $a_{\rm osc} = \sqrt{\hbar/m \omega_A}$ is the harmonic oscillator length. 
For convenience, in the following, we will drop the prime sign and assume that the trapping frequency in the transverse $x$-$y$ plane satisfies $\omega_A = \omega_B = \omega$. 
The quasi-2D harmonic trap is achieved by applying a much stronger trapping in the axial $z$-direction compared to that along the $x$-$y$ plane, i.e. $\omega_z \gg \omega$. 
Therefore, under the condition $\lambda = (\omega_z/\omega) \gg 1$, the wavefunction of each species can be factorized as follows 
\begin{equation}\label{2}
\Psi_{j} (x, y, z, t) = \psi_{j} (x, y, t) \phi_{j} (z),
\end{equation} 
where $\phi_{j} (z) $ is the normalized ground state wavefunction in the $z$-direction. 
Subsequently, the dimensionless form of the coupled GP equations after integrating 
over $\phi_{j}(z)$, results in the 2D form:
\begin{equation}\label{GP_new1}
\begin{split}
i\frac{\partial \psi_A(x,y,t)}{\partial t} = &
\bigg[- \frac{1}{2 } \nabla^{2}_{\perp}+ \frac{1}{2}(x^2 + y^2) \\ & + \sum_{j}^{} \mathcal{G}_{Aj} |\psi_j(x, y)|^2 \bigg] \psi_A(x,y,t),
\end{split} 
\end{equation}
and 
\begin{equation}\label{GP_new2}
\begin{split}
i\frac{\partial \psi_B(x,y,t)}{\partial t} = &
\bigg[- \frac{m_r}{2 } \nabla^{2}_{\perp}+ \frac{1}{2m_r} (x^2 + y^2) \\ & + \sum_{j}^{} \mathcal{G}_{Bj} |\psi_j(x, y)|^2 \bigg] \psi_B(x,y,t).
\end{split} 
\end{equation}
In these expressions, $\nabla_{\perp} ^ 2 = \partial^2_x + \partial^2_y$ connects to the kinetic energy term and $m_r = m_A/m_B$. 
Furthermore, $\mathcal{G}_{AA} = 2N_A \sqrt{2 \pi \lambda} a_{AA}/a_{\rm osc}$ and $\mathcal{G}_{BB} =  2m_rN_{B}\sqrt{2 \pi \lambda}a_{BB}/a_{\rm osc}$ refer to the intraspecies interaction strengths, while $\mathcal{G}_{jj^{'}} = N_{j^{'}} (1 + m_r)\sqrt{2 \pi \lambda} a_{jj^{'}}/a_{\rm osc}$ is the interspecies interaction strength. 

Regarding our mean-field calculations presented in the main text, {we numerically solve the above-described GP equations [Eqs.~\eqref{GP_new1},~\eqref{GP_new2}] using a split-time Crank-Nicolson method} adapted for binary condensates~\cite{Muruganandam2009, VUDRA_2021}. 
The initial ground state of the binary system is obtained by propagating the relevant equations in imaginary-time, until the solution converges to the desired state. 
Furthermore, the normalization of the $j$ species wavefunction is ensured by utilizing the transformation $\psi_j \rightarrow \frac{\psi_j}{\norm{\psi_j}}$ at every time-instant of the imaginary-time propagation until the energy of the desired configuration is reached with a precision $10^{-8}$. 
Having these solutions at hand as initial conditions, at $t= 0$, we study their evolution in real-time. 
The corresponding simulations are performed within a square grid containing $400 \times 400$ grid points with a grid spacing $\Delta x = \Delta y = 0.05$. 
The time-step of the integration $\Delta t$ is chosen to be $10^{-4}$.

\section{Dynamics after a modulation of the external confinement} 

\begin{figure*}[ht]
	\begin{center} 
		\includegraphics[height = 0.24\textheight, width = \linewidth]{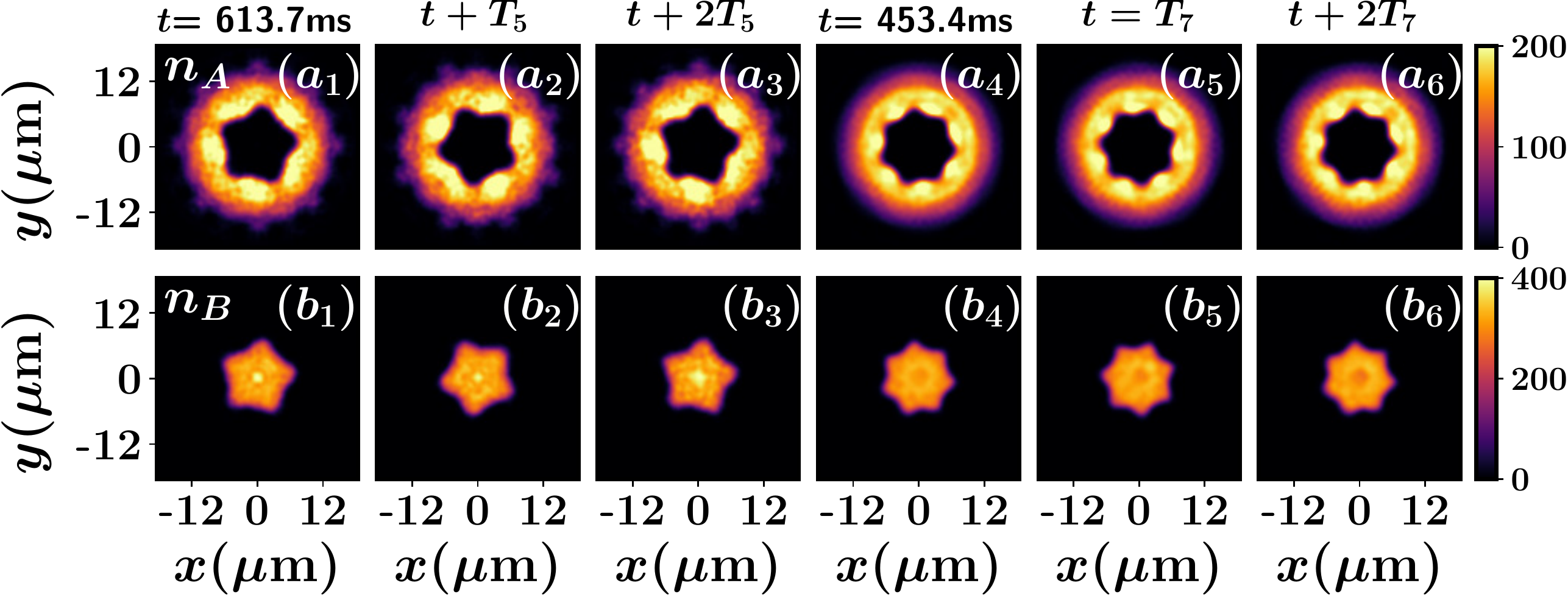}
		\caption{(Color online) Density snapshots of ($a_1$)-($a_6$) species A and ($b_1$)-($b_6$) species B for various symmetric patterns $D_5$ [($a_1$), ($a_2$), ($a_3$) ($b_1$), ($b_2$), ($b_3$)] and  $D_7$ [($a_4$), ($a_5$), ($a_6$), ($b_4$), ($b_5$), ($b_6$)] at selected time-instants (see legends) in the long-time dynamics. 
			$T_m$ is the driving period corresponding to the $m$-fold symmetric pattern $D_m$. 
			To induce the dynamics, the trapping potential of species B is modulated with amplitude $b = 0.24$ and frequency $\omega_D =2 \pi/T_m$. 
			The $^{87}$Rb-$^{85}$Rb binary BEC composed by $N_A=10^4$ and $N_B=5\times 10^3$ atoms is initialized in its ground state with $a_{AA}=99 a_B$, $a_{BB}=75a_B$, $a_{AB}=213a_{B}$ and trap frequency $\omega_A=\omega_B=2\pi \times 15$ Hz.}
		\label{fig5}
	\end{center}
\end{figure*}

In the main text, for the convenience of the Floquet analysis we assumed that the modulation of the scattering length of the core component has the same impact on the interface dynamics of the binary BEC as the one when applying a modulation of the external confinement of the core component. 
This is due to the fact that the natural angular frequencies $\omega_m$ [see below Eq.~(\ref{mathieu})] of the patterns do not depend on whether one of the systems' scattering lengths [see also the discussion in the next section] or the external trap frequency [see the description below] is being modulated as long as the corresponding modulation amplitude is small compared to the original value of the perturbed parameter. Furthermore, as can also be verified within the full GP calculations presented in the main text, the modulation dominantly affects the interface of the binary BEC where the species densities are vanishing since we operate in the deep immiscible interaction regime. 
Thus, in principle, it is possible to adjust the modulation amplitudes of the different time-dependent protocols such that the latter produce the same impact on the BEC interface. 
Along these lines, it is possible to equate the corresponding modulation terms at the interface, namely  
$\mathcal{A}\cos(\omega_Dt) \abs{\psi_B}^2 \psi_B = (1/2) m_B \omega^2 R^2 b \cos(\omega_D t) \psi_B$ which allow us to establish the relation $b = 2\mathcal{A} \abs{\psi_B}^2/(m_B \omega^2 R^2)$, where $\psi_B(r = R)$ is taken from the initial state obtained within the GP framework. 
Indeed, the latter formula connects the modulation amplitudes of the two dynamical protocols. 

To substantiate our above-mentioned argument regarding the same impact of the protocols on the interface, we subsequently solve the 2D GP equations of motion following a modulation of the confinement of the core component while keeping $a_{BB}$ fixed. 
In this case, the corresponding coupled system of GP equations reads 
\begin{equation}\label{GP_conf1}
\begin{split}
i\frac{\partial \psi_A(x,y,t)}{\partial t} = &
\bigg[- \frac{1}{2 } \nabla^{2}_{\perp}+ \frac{1}{2}(x^2 + y^2) \\ & + \sum_{j}^{} \mathcal{G}_{Aj} |\psi_j(x, y)|^2 \bigg] \psi_A(x,y,t),
\end{split} 
\end{equation}
and 
\begin{equation}\label{GP_conf2}
\begin{split}
& i\frac{\partial \psi_B(x,y,t)}{\partial t} =
\bigg[- \frac{m_r}{2 } \nabla^{2}_{\perp}+ \frac{1}{2m_r} (x^2 + y^2)\\ & \times(1 + b\cos(\omega_{D} t))  + \sum_{j}^{} \mathcal{G}_{Bj} |\psi_j(x, y)|^2 \bigg] \psi_B(x,y,t).
\end{split} 
\end{equation}

To demonstrate the connection with the results presented in the main text we consider a $^{87}$Rb-$^{85}$Rb binary BEC with $^{87}$Rb ($^{85}$Rb) being referred to as species A (B) in the following. 
Moreover, both species of the system are confined in a 2D harmonic trap with frequencies $\omega_A/2 \pi = \omega_B/2\pi \equiv \omega/2 \pi=15$ Hz. 
The intra- and interspecies scattering lengths are chosen to be $a_{BB} = 75a_B$, $a_{AA} = 99 a_B$ and $a_{AB} = 213 a_B$, while each species contains a particle number $N_A = 10^4$ and $N_B = 5 \times 10^3$ respectively. 
Consequently, the binary BEC is initially prepared in its ground state which corresponds to the phase-separated state described by the densities shown in Figs.~\ref{fig2} ($b$), ($c$) of the main text. 
To induce the dynamics we impose a periodic driving on the trapping potential of the core component as described in Eq. (\ref{GP_conf2}) with amplitude $b$ and frequency $\omega_D$. 
Note that the value $b=0.24$ is considered herein, which exactly corresponds to the modulation amplitude $\mathcal{A}=15 a_B$ of the scattering length of the core component examined in the main text, since $b = 2\mathcal{A} \abs{\psi_B}^2/(m_B \omega^2 R^2)$. 

Figure~\ref{fig5} depicts some representative density profiles of each species at specific time-instants of the long-time dynamics following the above-mentioned driving protocol on the confinement of the core component. 
As it can be readily seen, different symmetric patterns $D_m$ characterized by a respective $m$-fold symmetry build upon the densities of the individual species after every period of the modulation. 
For instance, we observe that five [Figs.~\ref{fig5}($a_1$)-($a_3$) and \ref{fig5}($b_1$)-($b_3$)] and seven [Figs.~\ref{fig5}($a_4$)-($a_6$) and \ref{fig5}($b_4$)-($b_6$)] fold star-shaped patterns are generated for  driving frequencies $\omega_D= 96$ Hz  and $\omega_D=165$ Hz respectively. 
Remarkably enough, these frequencies lie inside the corresponding resonant tongues obtained within Floquet theory and illustrated in Fig.~\ref{fig4} of the main text. 
Importantly, these $m$-fold star-shaped patterns patterns repeat themselves after a time $t' =t+ 2T_m$ and undergo a $\pi/m$ rotation every $t=T_m$. 
This behavior essentially manifests the sub-harmonic feature of the formed patterns, a phenomenon that has also been observed upon applying a modulation of the core component scattering length discussed in the main text. 
We also remark that also every other $m$-fold pattern can be created using different driving frequencies of the external confinement of the core component (results not shown).

Summarizing, we can deduce that the overall phenomenology observed in the interfacial dynamics of the binary BEC is the same when following a periodic modulation of either the scattering or the confinement of the core component. As a result, the assumptions made within the Floquet analysis are reasonably justified, at least, for the calculation of the natural frequencies of the generated patterns.

\section{Dynamical emergence of patterns by modulating the scattering length of the shell condensate}

In the main text, the pattern formation on the immiscible BEC interface has been demonstrated by applying a periodic modulation of the scattering length $a_{BB}$ of the core component consisting of $^{85}$Rb atoms. 
This particular parameter choice is especially motivated by the already demonstrated experimental feasibility to tune the scattering length of $^{85}$Rb atoms by means of Feshbach resonances~\cite{Papp}. 
In the following, we shall argue that the above-described parametric instability phenomenon, being an azimuthal symmetry breaking phenomenon, is quite generic for an immiscible condensate in the sense that is manifested through the periodic driving of any of the involved intra- and interspecies scattering lengths. 
\begin{figure*}[ht]
\begin{center} 
\includegraphics[height = 0.24\textheight, width = \linewidth]{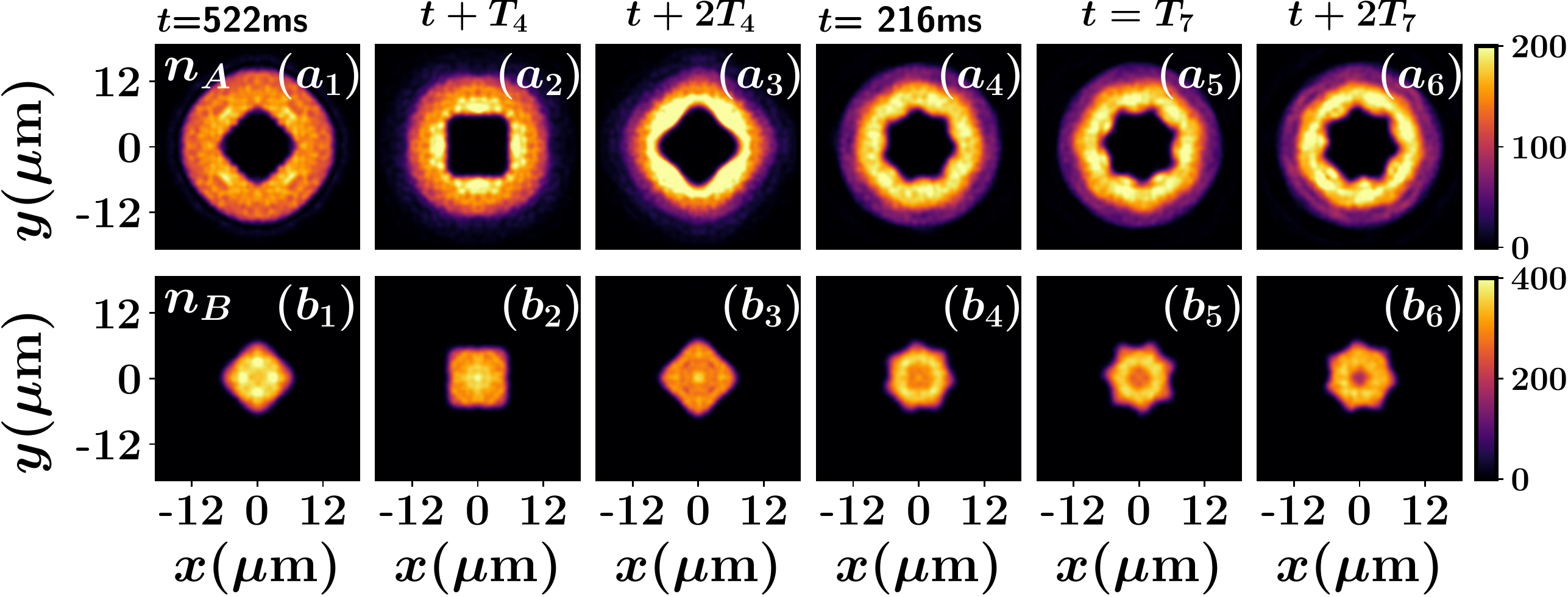}
\caption{(Color online) Density profiles of ($a_1$)-($a_6$) species A ($^{87}$Rb) and ($b_1$)-($b_6$) species B ($^{85}$Rb) exhibiting different symmetric patterns, i.e. $D_4$ [($a_1$), ($a_2$), ($a_3$) ($b_1$), ($b_2$), ($b_3$)] and $D_7$ [($a_4$), ($a_5$), ($a_6$), ($b_4$), ($b_5$), ($b_6$)] at selected time-instants (see legends) of the long-time dynamics. 
$T_m$ refers to the driving period at which the $m$-fold symmetric pattern $D_m$ dynamically appears. 
The dynamics is triggered by a periodic modulation of the $^{87}$Rb scattering length
with amplitude $\mathcal{A} = 20 a_B$ and frequency $\omega_D = 2 \pi/T_m$.  
The $^{87}$Rb-$^{85}$Rb binary BEC possesses $N_A = 10^4$ and $N_B = 5\times 10^3$ atoms and it is initialized in its ground state with $a_{AA}= 99 a_B$, $a_{BB}= 75a_B$, $a_{AB}= 213a_{B}$ and trapping frequency $\omega_A=\omega_B=2\pi \times 15$ Hz. }
\label{g_AA}
\end{center}
\end{figure*} 

To support our arguments we demonstrate that it is possible to dynamically generate the above-mentioned patterns following a time periodic modulation of the {\it shell} condensate scattering length $a_{AA}$. 
More specifically, we consider a binary bosonic mixture of $^{87}$Rb-$^{85}$Rb atoms, labeling $^{87}$Rb [$^{85}$Rb] as species A [B]. 
Both species are confined in a 2D isotropic harmonic trap with frequencies $\omega_A= \omega_B \equiv \omega=2 \pi \times 15$ Hz. 
The intra- and interspecies scattering lengths are chosen to be $a_{BB} = 75a_B$, $a_{AA} = 99 a_B$ and $a_{AB} = 213 a_B$ while each species contains a particle number $N_A = 10^4$ and $N_B = 5 \times 10^3$. Evidently, the initial state is the same as the one considered in the main text, namely the $^{87}$Rb and the $^{85}$Rb form the core and shell condensates respectively. 
To trigger the dynamics, we apply the following modulation $\tilde{a}_{AA}
= a_{AA} + \mathcal{A}\cos(\omega_D t)$ of the scattering length of the shell condensate. 
In particular, we consider $\mathcal{A} = 20a_B$ such that $\mathcal{A}/a_{AA} \approx 0.2$. 

Characteristic density profiles of each species in the course of the time-evolution are presented in Fig.~\ref{g_AA} where the pattern formation occuring at the interface of the binary BEC is illustrated. 
In particular, as a case example, we showcase the dynamical formation of four fold [Figs.~\ref{g_AA}($a_1$)-($a_3$) and Figs.~\ref{g_AA}($b_1$)-($b_3$)] and seven fold [Figs.~\ref{g_AA}($a_4$)-($a_6$) and Figs.~\ref{g_AA}($b_4$)-($b_6$)] symmetric patterns following the periodic driving of the $^{87}$Rb scattering length. 
These patterns are generated sub-harmonically, which is evident from their repetition at $t' = t + 2 T_{m}$.
The above-mentioned patterns are realized for driving frequencies $\omega_D= 69$ Hz ($D_4$) and $\omega_D= 168$ Hz ($D_7$) respectively. 
We remark that also other patterns possessing a different $m$-fold symmetry can be formed within this protocol in the considered setting (results not shown for brevity). 
Finally, we mention in passing that the same overall phenomenology can be identified by considering the same setup (as in the main text) but modulating the interspecies scattering length $a_{AB}$ namely $\tilde{a}_{AB} = a_{AB} + \mathcal{A}\cos(\omega_D t) $ e.g. with $\mathcal{A} = 40 a_B$. 
Remarkably, also in this case a variety of $m$-fold symmetric patterns can again be realized for distinct driving frequencies $\omega_D$ (results not shown).       

\section{Pattern formation in a strongly mass-imbalanced mixture} 

To expose the general character of our findings regarding the emergent star-shaped patterns building upon the interface of immiscible binary BECs, we next demonstrate as a case example the dynamics of the experimentally relevant strongly mass-imbalanced $^{41}$K-$^{87}$Rb mixture \cite{modugno2002two}. 
Before proceeding, it is also worth mentioning that the same pattern formation can also be generated in completely mass-balanced bosonic mixtures, e.g. by considering two hyperfine states of $^{87}$Rb (not shown for brevity). 
Moreover, we have also performed the modulation dynamics using the following particle number ratio $N_B/N_A = 0.7, 0.85,1$, trapping frequencies $10$ Hz and $25$ Hz and found that also these systems show a similar overall phenomenology to the one discussed in the main text. 

For simplicity, in the following we label $^{41}$K as species A and $^{87}$Rb as species B with $m_r = 0.47$ and particle number in each component $N_A=5\times 10^3$ and $N_B=5\times 10^3$. 
Moreover, we employ the experimentally realizable (for this mixture) values of the intra- and interspecies scattering lengths \cite{marte2002feshbach,wang2000ground,ferlaino2006feshbach}. 
These correspond to $a_{AA}= 65a_{B}$ and $a_{BB}= 99a_{B}$ for the $^{41}$K and $^{87}$Rb atoms respectively, whilst the interspecies one is fixed to the value $a_{AB}=163a_{B}$. 
Also, we use a 2D harmonic oscillator potential of frequency $\omega_A = \omega_B = 2 \pi \times 15$ Hz. 
As before, the system is initially prepared in its immiscible ground state where now the $^{87}$Rb atoms (heavier species) configure the core condensate and the $^{41}$K atoms (lighter species) form a shell around it. 
To trigger the dynamics, we subsequently modulate the scattering length $a_{BB}$ time-periodically with amplitude $\mathcal{A}$ and frequency $\omega_D$ according to the protocol $\tilde{a}_{BB} = a_{BB} + \mathcal{A}\cos(\omega_D t)$. 
We also consider $\mathcal{A} = 20 a_{B}$ such that the ratio $\mathcal{A}/a_{BB} \approx 0.2$ is the same as in the main text.

\begin{figure*}[ht]
	\begin{center} 
		\includegraphics[height = 0.24\textheight, width = \linewidth]{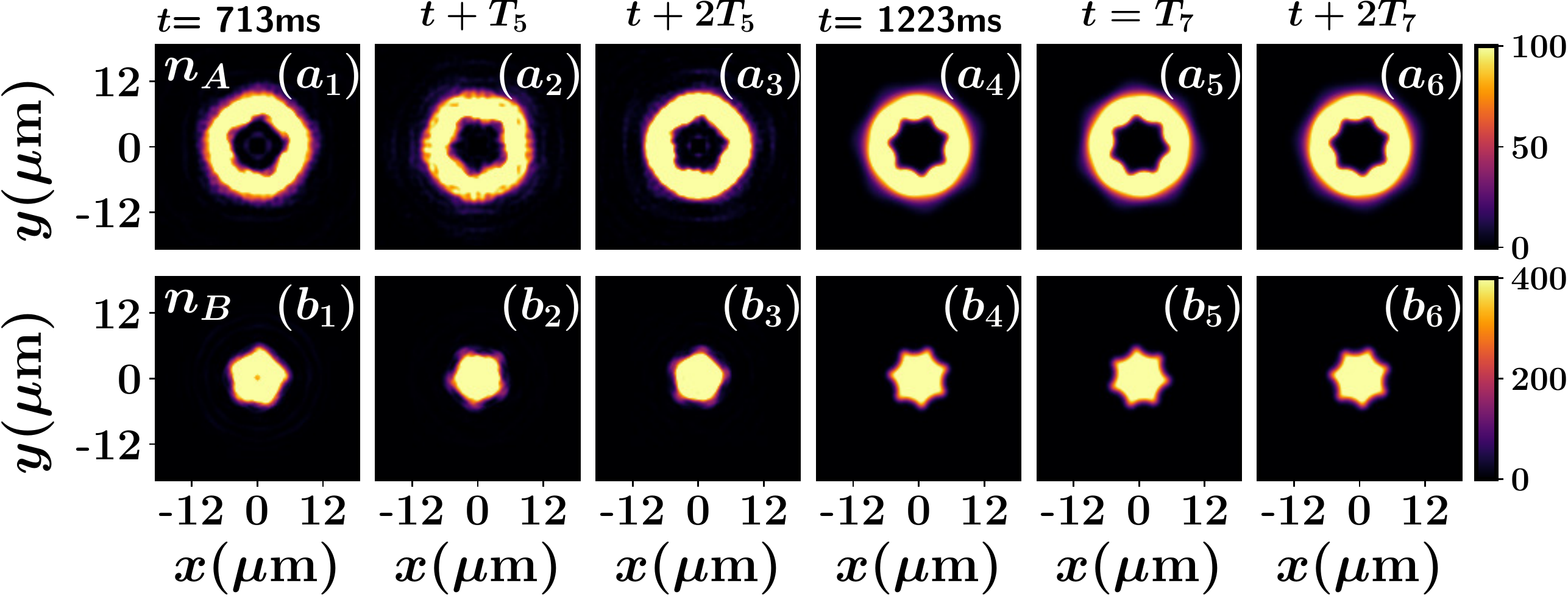}
		\caption{(Color online) Density profiles of ($a_1$)-($a_6$) species A ($^{41}$K) and ($b_1$)-($b_6$) species B ($^{87}$Rb) showing different symmetric patterns namely $D_5$ [($a_1$), ($a_2$), ($a_3$) ($b_1$), ($b_2$), ($b_3$)] and $D_7$ [($a_4$), ($a_5$), ($a_6$), ($b_4$), ($b_5$), ($b_6$)] at selected time-instants (see legends) of the long-time dynamics. 
			$T_m$ refers to the driving period corresponding to the $m$-fold symmetric pattern $D_m$. 
			The dynamics is induced by a periodic modulation of the $^{87}$Rb scattering length
			with amplitude $\mathcal{A} = 20 a_B$ and frequency $\omega_D = 2 \pi/T_m$.  
			The $^{41}$K-$^{87}$Rb binary BEC contains $N_A = 5\times 10^3$ and $N_B = 5\times 10^3$ atoms and it is prepared in its ground state with $a_{AA}=65 a_B$, $a_{BB}=99a_B$, $a_{AB}= 163a_{B}$ and trapping frequency $\omega_A=\omega_B=2\pi \times 15$ Hz. }
		\label{rbk}
	\end{center}
\end{figure*}

Characteristic density profiles of each species showcasing five fold [Figs.~\ref{rbk}($a_1$)-($a_3$) and Figs.~\ref{rbk}($b_1$)-($b_3$)] and seven fold [Figs.~\ref{rbk}($a_4$)-($a_6$) and Figs.~\ref{rbk}($b_4$)-($b_6$)] symmetric patterns are presented in Fig.~\ref{rbk} after applying the periodic driving of the $^{87}$Rb scattering length. 
More specifically, the above-mentioned patterns are realized for driving frequencies $\omega_D=73$ Hz ($D_5$) and $\omega_D=98$ Hz ($D_7$) respectively. 
Of course, it is possible to dynamically create all the different $m$-fold patterns by using the appropriate driving frequencies (not shown for brevity). 
Remarkably, the patterns appearing in Fig. \ref{rbk} feature a similar to the previously discussed dynamics, namely a particular $m$-fold symmetric pattern $D_m$ is repeated exactly at the same location at $t+2 T_m$ (where $T_m$ is the time period of the driving) thus revealing its sub-harmonic nature. 
However, due to the significant mass-imbalance the same patterns are realized at a different driving frequency when compared to the sightly mass-imbalanced $^{87}$Rb-$^{85}$Rb scenario. 
This behavior is also supported by the $\omega_m (m)$ dispersion relation derived within the Floquet theory in the main text [see also below Eq.~\eqref{mathieu}]. 
Summarizing, according to our mean-field calculations we can deduce that the significant mass-imbalance between the species affects the natural angular frequencies of the emergent patterns and the timescale of the appearance of the relevant phenomenology.

\section{Derivation of the Mathieu Equation}

Here we discuss in detail the derivation of the Mathieu equation within the Floquet analysis \cite{Staliunas2002,goldman2014periodically,barone1977floquet,eckardt2017colloquium} presented in the main text. 
In particular, we start from a pair of 3D equations which are subsequently reduced to a form including only a single degree-of-freedom that is able to describe the emergent pattern formation in 2D. 
Following the Madelung transformation~\cite{madelung1927quantentheorie} of the condensate wavefunction namely $\psi_j(r, \theta, z) = \sqrt{n_j(r, \theta, z)}e^{i \phi }$, with $n_j$, $\phi_j$ being the density and the phase of the $j$ species and using the superfluid velocity $\boldsymbol{v}_j = \frac{\hbar}{m_j}\nabla \phi_j$~\cite{pethick_smith_2008}, the hydrodynamic form of the GP equations regarding $\boldsymbol{v}_j$ take the form  
\begin{equation}\label{eq1}
	- m_A \frac{\partial \boldsymbol{v}_A}{\partial t} = \frac{\vb{\nabla} P_A}{n_A},
~~~~	- m_B \frac{\partial \boldsymbol{v}_B}{\partial t} = \frac{\vb{\nabla} P_B}{n_B}.
\end{equation}
Integrating these equations we can easily show that the effective pressure terms of the species $P_j$ satisfy   
\begin{equation}\label{pressure_diff}
P_B - P_A =  \hbar n_A \frac{\partial \phi_A}{\partial t}- \hbar n_B \frac{\partial \phi_B}{\partial t}+C,
\end{equation} 
where $C$ is an integration constant.
Subsequently, we decompose the fluid pressure $P_j$ into a so-called static part, $P^{\rm s}_{j}$, and a dynamical one $P^{\rm d}_{j}$. 
In particular, before the onset of the patterns only the static pressure is present in the system, while both of them exist afterwards. 
Note that $P_j^s$ can be obtained considering that the phase $\phi_j$ remains constant before the undulation of the interface starts, i.e. $\frac{\partial \phi_j}{ \partial t } \approx 0$. 
Therefore considering that $P_A = \frac{1}{2}(m_A n_A v^2_A) + \frac{\hbar^2 \sqrt{ n_A}}{2 m_A } \nabla^2 \sqrt{n_A} + g_{AA}n^2_A + \frac{1}{2}m_A n^2_A \omega^2 ( r^2 + \lambda_A z^2 )$ and $P_B = \frac{1}{2}(m_B n_B v^2_B) + \frac{\hbar^2 \sqrt{n_B}}{2 m_B } \nabla^2 \sqrt{n_B} + g_{BB}n^2_B + \frac{1}{2}m_B n_B \omega^2 ( r^2
+ \lambda_B z^2 )1 + \frac{1}{2}m_B n_B \omega^2 r^2 b \cos(\omega_D t)$ it holds that
\begin{equation}
\begin{split}
 P^{\rm s}_{B} - P^{\rm s}_A= &\frac{1}{2} m_B \omega^2 n_Br^{2}\cos(\omega_D t) \\& + g_{BB}n^2_{B} -g_{AA}n^2_{A} + C.
\end{split}
 \end{equation} 
To arrive at the last equation we have assumed that the background density is uniform and as a result the modulation of the harmonic potential impacts only the BEC interface. 
The constant $C' = g_{BB}n^2_{B} - g_{AA}n^2_{A} + C$ can be found from the standard pressure jump condition at the interface [see also Eq.~\eqref{SB}]. 
For a cylindrical surface, which we have considered herein, $R_2 \rightarrow \infty $ and $R_1 \rightarrow R$. 
As a consequence
\begin{equation}
\begin{split}
	\big [ P^{\rm s}_B- P^{\rm s}_A \big ]_{r = R} = \frac{1}{2}m_B &\omega^2 n_{B} b R^2 \cos(\omega_D t) \\& + C^{'} =  \frac{\sigma}{R}.
\end{split}
\end{equation} 
which readily implies that 
\begin{equation}
C' = -\frac{1}{2}m \omega_B^2n_B b R^2 \cos(\omega_Dt) + \frac{\sigma}{R}.
\end{equation}

In this way, the difference of the static pressure term between the species is given by
\begin{equation}
	P^{\rm s}_B - P^{\rm s}_A = \frac{1}{2}m_B\omega^2 n_Bb(r^2 - R^2)\cos(\omega_D t) + \frac{\sigma}{R},
\end{equation}
and accordingly the dynamical one acquires the form
\begin{equation}
	P^{\rm d}_B - P^{\rm d}_A = \hbar n_A \frac{\partial \phi_A}{\partial t}- \hbar n_B \frac{\partial \phi_B}{\partial t}.
\end{equation}

Furthermore, after the onset of the instability, the curved surface of the cylinder is deformed. 
Accordingly, for a deformed cylindrical surface with deformation $\zeta$, the right-hand-side of 
Eq.~\eqref{SB} can be linearized~\cite{Lamb_1932} as follows
\begin{equation}\label{f_s_r}
\sigma \big[ \frac{1}{R_1} + \frac{1}{R_2}\big] = \frac{\sigma}{R} - \sigma[\frac{1}{R^2} + \frac{1}{R^2}\frac{\partial^2}{\partial\theta^2} + \frac{\partial^2}{\partial z^2} ] \zeta.
\end{equation}
Moreover, the left-hand-side of Eq.~\eqref{SB} can be linearized  around $r  = R$
(using the Taylor series expansion) as 
\begin{equation}\label{f_s_l}
\begin{split}
	 [P_B - P_A]_{r = R + \zeta} = & \frac{\sigma}{R} + n_B m_B \omega^2 b R^2 \cos(\omega_D t) \zeta \\ & 
	+ \hbar n_A \frac{\partial \phi_A}{\partial t} - \hbar n_B \frac{\partial \phi_B}{\partial t}.
\end{split}
\end{equation} 
Equating the above Eq.~\eqref{f_s_r} and Eq.~\eqref{f_s_l} we then get
\begin{equation}
\begin{split}\label{main}
	\hbar n_A \frac{\partial \phi_A}{\partial t} - \hbar n_B \frac{\partial \phi_B}{\partial t} = & -m_B \omega^2 n_B b R^2 \cos(\omega_D t) \zeta \\ & -
	\sigma \big[\frac{1}{R^2} + \frac{1}{R^2}\frac{\partial^2}{\partial\theta^2} + \frac{\partial^2}{\partial z^2} \big ] \zeta.
	\end{split}
\end{equation}
Let us then write the deformation $\zeta(r, \theta, z, t)$ in the form 
\begin{equation}\label{expansions_zeta}
\zeta(\theta,z,t)=\sum_{m=1}^{\infty}{\zeta}_m(t) e^{i({m\theta+kz})}.
\end{equation} 
Furthermore, the phase terms $\phi_A$, $\phi_B$ following the solution of the Laplace equation \cite{jackson2007classical} 
$\nabla^2\phi_j = 0$ can be expressed as 
\begin{equation}\label{expansions_phi_1}
\phi_A(r,\theta,z,t)=\sum_{m=1}^{\infty}  \mathcal{P} K_m(kr) e^{i({m\theta+kz})},
\end{equation}
and
\begin{equation}\label{expansions_phi_2}
\phi_B(r,\theta,z,t)=\sum_{m=1}^{\infty}\mathcal{Q} I_m(kr) e^{i({m\theta+kz})},
\end{equation}
where $I_m(kr)$ and $K_m(kr)$ denote the $m^{th}$ order modified Bessel functions of the first- and second-kind respectively \cite{abramowitz1964handbook}. 
Also, $\mathcal{P}$, $\mathcal{Q}$ are constants while the integers $m$, $k$ are the azimuthal and axial wavenumbers respectively.

The kinematic boundary condition is defined as~\cite{Lamb_1932}
\begin{equation}\label{kbc}
\frac{\partial \zeta}{\partial t}=v_{Ar}(r=R)=v_{Br}(r=R).
\end{equation}
Employing the Eq.~\eqref{kbc}we arrive at 
\begin{equation}\label{A}
\begin{split}
\frac{d{\zeta}_m(t)}{dt}= \frac{\hbar k}{m_A}\mathcal{P} K'_m(kR)=
\frac{\hbar k}{m_B}\mathcal{Q} I'_m(kR).
\end{split}
\end{equation}
Consequently, $\mathcal{P}$ and $\mathcal{Q}$ can be obtained from Eqs.~\eqref{A}. 
Then, by utilizing Eqs.~\eqref{expansions_phi_1} and \eqref{expansions_phi_2}, we can write the 
phase of the A species as follows 
\begin{equation}\label{expansions_phi_v1}
\phi_A(r,\theta,z,t)=\sum_{m=1}^{\infty}\frac{d{\zeta}_m(t)}{dt}  \frac{m_A K_m(kr)}{\hbar kK'_m(kR)} e^{i({m\theta+kz})},
\end{equation}
and for the B species 
\begin{equation}\label{expansions_phi_v2}
\phi_B(r,\theta,z,t)=\sum_{m=1}^{\infty}\frac{d{\zeta}_m(t)}{dt}  \frac{m_B I_m(kr)}{\hbar kI'_m(kR)} e^{i({m\theta+kz})}.
\end{equation}
Substituting Eqs.~\eqref{expansions_phi_v1} and \eqref{expansions_phi_v2} into Eq.~\eqref{main}, we get
\begin{equation}
\begin{split}
& \sum_{m = 1}^{\infty} \bigg[	n_A m_A\frac{K_m(kr)}{kK^{'}_m(kR)}\frac{d^2 \zeta_m}{d t^2} - n_B m_B \frac{I_m(kr)}{kI^{'}_m(kR)}\\& \times \frac{d^2 \zeta_m}{d t^2} \bigg]e^{i(m\theta +k z)}
= \sum_{m = 1}^{\infty} \bigg[- m_B \omega^2 R n_{B} b\zeta_m \\& \times \cos(\omega_Dt) +\sigma \bigg(\frac{m^2 - 1}{R^2} + k^2 \bigg) \bigg] \zeta_m e^{i (m\theta +k z)}.
\end{split}
\end{equation}
Next we let $k \rightarrow 0$ and use $\lim\limits_{k \rightarrow 0}\frac{K_m(kr)}{kK^{'}_m(kR)} = \lim\limits_{k \rightarrow 0}\frac{I_m(kR)}{kI^{'}_m(kR)}= \frac{R}{m}$ to arive at
\begin{equation}\label{Mathieu1} 
\begin{split}
\frac{d^2 \zeta_m}{d t^2} + \frac{\sigma m(m^2 -1)}{R^3(m_B n_B - m_A n_A)} & \bigg[1 - \frac{m_B \omega^2 n_B R^3b}{\sigma(m^2  - 1)} \\& \times \cos(\omega_D t) \bigg ] \zeta_m = 0.
\end{split}
\end{equation}
Eq.~\eqref{Mathieu1} has the form of the so-called Mathieu equation used in the main text. 
Note that Eq.~\eqref{Mathieu1} possesses only one degree-of-freedom i.e. $\zeta_m$ which is the amplitude (associated with the radial direction) of the $m$-th mode (related to the azimuthal direction), thus highlighting the 2D nature of the patterns. 
Moreover, the choice of $k \rightarrow 0$ indicates the absence of wave excitations along the $z$-direction. 
The latter ensures that the dynamics of the system is "frozen" in the $z$-direction. 
Recall that we have also utilized this approximation in order to reduce the full 3D GP equations of motion into the 2D ones. 
Another important observation is that the parameters $R$, $n_B$ and $n_A$ determine the natural angular frequencies of the emergent patterns, whilst the values of these parameters used within the Floquet analysis are taken from the initial state obtained via the full GP calculations. 
Hence even though the interspecies interaction $g_{AB}$ does not explicitly appear in the relevant equations of the Floquet analysis, its effect is implicitly included in the values of $R$, $n_B$ and $n_A$. 
In other words, any modification of $g_{AB}$ in the initial state of the binary BEC would definitely shift the natural angular frequencies of the patterns, since it alters the magnitude of $R$, $n_B$ and $n_A$. 
For instance, if $g_{AB}$ is increased then $R$ decreases and consequently $n_B$ increases leading in turn to a modification of the respective natural angular frequency $\omega_m$ [see also below Eq.~(\ref{mathieu}) in the main text].

\end{document}